\documentclass[aps,prl,twocolumn, nofootinbib]{revtex4}
\usepackage{bm}
\usepackage{latexsym}
\usepackage{amsmath,amsfonts,amssymb}
\usepackage{graphicx,epsfig}
\usepackage[utf8]{inputenc}
\usepackage{array}
\usepackage{wrapfig}
\usepackage{multirow}
\usepackage{tabularx}
\usepackage{psfrag}
\usepackage{subfigure}
\usepackage{amsthm}
\usepackage{color}

\def\be{\begin{equation}}
\def\ee{\end{equation}}
\def\ber{\begin{eqnarray}}
\def\eer{\end{eqnarray}}
\def\tred{\textcolor{red}}

\def\D{\Delta}
\usepackage{amsmath}
\usepackage{braket}
\usepackage{indent first}
\usepackage[normalem]{ulem}
\useunder{\uline}{\ul}{}

\begin{document}
\title{
A novel mechanism for probing the Planck scale}

\author{Saurya Das}
\email{saurya.das@uleth.ca}
\affiliation{Theoretical Physics Group and Quantum Alberta, Department of Physics and Astronomy, University of Lethbridge, 4401 University Drive, Lethbridge, Alberta T1K 3M4, Canada}

\author{Sujoy K. Modak}
\email{smodak@ucol.mx}
\affiliation{Facultad de Ciencias - CUICBAS, Universidad de Colima, Colima, C.P. 28045, M\'exico}

\begin{abstract}
{The Planck or the quantum gravity scale, being $16$ orders of magnitude greater than the electroweak scale, is often considered inaccessible by current experimental techniques. 
However, it was shown recently by one of the current authors
that quantum gravity effects via the Generalized Uncertainty Principle affects the time required for free wavepackets to double their size, and this difference in time is at or near current experimental accuracies \cite{carlos1, carlos2}.
In this work, we make an important improvement over the earlier study, by taking into account the leading order relativistic correction, which naturally appears in the sytems under consideration , 
due to the significant mean velocity of the travelling wavepackets. 
Our analysis shows that although the relativistic correction adds nontrivial modifications to the results of \cite{carlos1, carlos2}, the earlier claims remain intact and are in fact strengthened. We explore the potential for these results being tested in the laboratory. }
%

\end{abstract}
\maketitle

%
The Planck length, energy and time scales, first  
mentioned by Planck himself in \cite{planck, tomil},
continues to play special roles in physics. 
While this are believed to be the scales where Quantum Gravity (QG) effects will most certainly appear, given the immense gap between the electroweak scale ($\simeq 1$ TeV)
and Planck scale ($\simeq 10^{16}$ TeV), 
it is conceivable that some of these effects may show up in this intermediate region, even if indirectly. It is also believed that the Planck scale signifies an
absolute minimum measurable length scale in Nature, beyond which the notion of a 
continuum spacetime seizes to exist. 
Arguments in favour of a Minimum Length scale (MLS) can also 
be found in early works of Heisenberg \cite{hberg}, Yang \cite{yang}, Deser \cite{deser} 
and Mead \cite{maed1, maed2}. They have been refined further in many recent works (see e.g. \cite{garay}).

Although the Planck scale, MLS and the QG scale are often 
 assumed to be of the same order of
 magnitude, {\it per se}, there is no reason or evidence behind this assumption. We will therefore relax this, and assume that new physical effects, including QG effects may potentially show up in the vast arena of 
$15$ orders of magnitude intervening between the electroweak and the Planck scales. Therefore, in the absence of a direct probe beyond the LHC scale energy ($\simeq 10$ TeV), it is imperative that one looks for potential experimental signatures and new physics that may be present in the aforementioned energy range. 


In this letter, we examine this idea and expand on the related work first proposed by one of us in \cite{carlos1,carlos2}, in which a concrete proposal was made to examine the hypothesized  fundamental minimal scale in Nature in an indirect manner.
The way it works is as follows: 
we know that wavepackets in quantum mechanics broaden in time as they evolve via a free Hamiltonian, and the rate of this broadening can be estimated accurately. In particular, it is straightforward to compute the time taken for wavepackets to double their size as they evolve via a free Hamiltonian. Width of wavepackets are often measured in Atomic-Molecular-Optical (AMO) experiments for various purposes ({\it {e.g.}} \cite{amo1}, \cite{amo2}). In this work, we re-examine this effect, but in light of a Hamiltonian which is still free, {\it but}
modified from the canonical Hamiltonian due to the
Generalized Uncertainty Principle (GUP), which encapsulates a MLS and is implied by it. 
Such a generic modification of the Heisenberg Uncertainty Principle (HUP) has been argued from many theories of QG, including 
String Theory, Loop Quantum Gravity, Doubly Special Relativity, black hole physics etc, and its implications were examined 
\cite{gup1,gup2,gup3,smolin,kpp,golam,gup4,dv1,adv,bdm,doug}.

Following earlier work by one of the current authors,  
in this paper, we examine promising experimental paths which might be able to detect GUP modifications 
with a high accuracy. In particular, the
``doubling time difference (DTD)'' 
(difference in times taken for a wavepacket to double in size, with and without GUP)
was computed in \cite{carlos1, carlos2}. It was also shown there however, that the DTD only becomes experimentally measurable, once the velocity of the travelling wave-packets are quite large ($\approx\,10^5 - 10^6$ m/s). This is because the GUP effects are momentum (and hence velocity) dependent and gets enhanced with increasing velocity of the wavepackets. While this is encouraging, one encounters the following issue: for these velocities, the 
relativistic corrections are of the order of 
$(v/c)^2 \simeq 10^{-6}-10^{-4}$, and it has to be 
determined whether these corrections will be comparable or
exceed the GUP corrections for the energy and momenta range under consideration. 
It is precisely this important point that we will examine in this paper and show that the GUP effects are still potentially measurable! 
In an attempt to systematically study both the relativistic and GUP effects, we in fact find that the two get mixed in a non-trivial way. However, it is still possible to appropriately `filter out' the relativistic effects and extract the GUP corrections, which are again just within the realm of current and future experimental acuracies.


We start by considering the Hamiltonian for a free particle of mass $m$ in $(1+1)$-dimensions, including the leading order relativistic correction term
\begin{eqnarray}
H
&=& \frac{p^2}{2m} - \frac{p^4}{8m^3c^2} \label{ham1} 
\end{eqnarray}
Now, as per GUP, the fundamental commutator between position and momentum is modified to 
\cite{adv}
\begin{eqnarray}
[x,p] = i\hbar\, [1-2\alpha p +4 \alpha^2p^2]~.
\end{eqnarray}
The above defines a minimum measurable length and a maximum measurable momentum, in terms of the GUP parameter $\alpha$ \cite{carlos1}
\begin{equation}
 (\Delta x)_{\text{min}} 
 = \frac{3\,\alpha_0}{2}\,\ell_{Pl};
 ~~~(\Delta p)_{\text{max}} 
 = 
 \frac{M_{Pl}\,c}{2 \alpha_0}
\label{alpha}
\end{equation}
where we have defined  
$\alpha= \alpha_0/M_{Pl} c$, $\alpha_0$ being dimensionless. 
$M_{Pl}$ is the Planck mass, 
$M_{Pl} c$ the Planck momentum, 
$M_{Pl} c^2 \approx 10^{16}$ TeV the Planck energy and 
$\ell_{Pl} \approx 10^{-35}$ m is the Planck length. 
We do not assume any specific value of $\alpha_0$, rather we hope that experiments will shed light on the allowed values of $\alpha_0$. 
Since no evidence of a MLS has not been found in experiments at the LHC, one is forced to put an upper bound on $\alpha_0$. Together with a lower bound on it corresponding to the Planck scale, 
one arrives at the following allowed range:
$1 \leq \alpha_0 \leq 10^{16}$.

 
%

Next, for calculational convenience, 
we define an auxiliary momentum variable 
$p_0$, which is `canonical' in the sense that $[x,p_0]=i\hbar$, and therefore as an operator, one can write $p_0=-i\hbar\,d/dx$. 
This is related to the physical (i.e. measurable) momentum $p$ via the relation $p= p_0(1-\alpha p_0 + 2 \alpha^2 p_0^2)$. Substituting in Eq.(\ref{ham1}), one obtains the following effective Hamiltonian for a relativistic system, incorporating GUP
{
\begin{eqnarray}
H
%
%
&=& H_{\text{NR}} + H_{\text{rel}} + H_{\text{LGUP}} +\nonumber \\ && + H_{\text{QGUP}} + H_{\text{LGUP}}^{\text{rel}}
\label{hfull}
\end{eqnarray}
where,
(i) $H_{\text{NR}} = \frac{p_0^2}{2m}$,
(ii) $~H_{\text{rel}} = -\frac{p_0^4}{8m^3c^2}$,
(iii) $~H_{\text{LGUP}} = - \frac{\alpha}{m} p_0^3,$
(iv) $~H_{\text{QGUP}} = \frac{5\alpha^2}{2m} p_0^4$,
and (v) $~H_{\text{LGUP}}^{\text{rel}} = \frac{\alpha}{2m^3c^2}p_0^5$. 
In the above, 
(i) is the standard non-relativistic Hamiltonian, 
(ii) the leading order relativistic correction, 
(iii) the linear GUP correction (proportional to $\alpha$), 
(iv) the quadratic GUP correction 
(proportional to $\alpha^2$)
and (v) the hybrid or mixed term,
which includes both the relativistic and linear GUP correction.  
%
%
%
%
 }

%
%

\tred{
}

Next, we move on to the study of evolution of free wavepackets under the above Hamiltonian. 
It is textbook knowledge that a free wave-packet tends to broaden itself due to the Heisenberg's uncertainty principle. Use of the Ehrenfest theorem is one of the direct ways of estimating this broadening. 
Here our interest is to consider the modified broadening rate of the free wave-packet with the full Hamiltonian \eqref{hfull}. 
As is well-known, the Ehrenfest's theorem gives the time derivative of the expectation values of the position 
($x$) and its canonically conjugate momentum ($p_0$) operators as follows:
$\frac{d}{dt} \langle x \rangle = \frac{1}{i \hbar} \langle [ x, H ] \rangle = \Big\langle \frac{\partial H}{\partial p_0} \Big\rangle$ 
and
 $\frac{d}{dt} \langle p_0 \rangle = \frac{1}{i \hbar} \langle [p_0, H] \rangle = - \Big\langle \frac{\partial H}{\partial x} \Big\rangle$. These can be extended to the expectation of any operator of course, and in particular to $p_0^n$, which appear in (\ref{hfull}) for various integer values of $n$. 
For the above, one obtains
 $\frac{d}{dt} \langle p_0^n \rangle = \frac{1}{i \hbar} \langle [ p_0^n, H ] \rangle =0$, implying that 
 $\langle p_0^n \rangle = \text{constant in time}$.
 
%


Next, to estimate the DTD, we first write the first and second time-derivatives of the 
square of the width (or variance) of the quantum mechanical wave-packet, which is 
defined as $\xi = \Delta x^2 = \langle x^2\rangle -\langle x \rangle^2$:
\ber
\dot{\xi} &=& \frac{d\xi}{dt} = \frac{d}{dt} \langle x^2 \rangle - 2 \langle x \rangle \frac{d  \langle x\rangle}{dt} \label{dxi}\\
\ddot{\xi} &=& \frac{d^2\xi}{dt^2} = \frac{d^2}{dt^2} \langle x^2\rangle - 2 \left(\frac{d  \langle x \rangle}{dt}\right)^2 - 2x \frac{d^2 \langle x \rangle}{dt^2}~. \label{ddxi}
\eer
The above can be simplified using the Ehrenfest theorem and the Hamiltonian given in (\ref{hfull}). 
%

To calculate the contributions for all the terms in
(\ref{hfull}), we consider each term in addition to the free nonrelativistic term ($p_0^2/2m$) separately, and write $H=\frac{p_0^2}{2m} + D p_0^{n}$ with $n>2$ and $D$ a constant, and compute the corresponding correction, using the 
Ehrenfest theorem and $[x,p_0]=i\hbar$. 
 Finally, we plug-in the appropriate value of $n$ and $D$ for each correction term in (\ref{hfull}) and add them together to find the total correction. 
%
 A straightforward calculation of the \eqref{dxi} and \eqref{ddxi} then yields,
\ber
\dot{\xi} &=& \frac{1}{m}\big(\langle  x p_0 + p_0 x\rangle - 2 \langle p_0 \rangle \langle x \rangle\big) + \nonumber \\
&& nD \big(\langle  x p_0^{n-1} + p_0^{n-1} x\rangle - 2 \langle p_0^{n-1} \rangle \langle x \rangle\big)  \\
\ddot{\xi} &=& \frac{2}{m^2} \Delta p_0^2 + \frac{4nD}{m} (\langle p_0^n\rangle - \langle p_0 \rangle \langle p_0^{n-1} \rangle) \nonumber \\
&&+ 2n^2 D^2 \Delta {p_0^{(n-1)}}^2 ~.
\eer
In the above, $\Delta p_0^2 = \langle p_0^2\rangle - \langle p_0 \rangle^2$ is the variance of 
the canonical momentum and $\Delta {p_0^{(n-1)}}^2 = \langle p_0^{2(n-1)} \rangle - \langle p_0^{n-1} \rangle^2$, that of the $(n-1)$-th power of the canonical momentum. We can now
identify $n$ and $D$ for all higher order corrections to the NR Hamiltonian and put them in the above expression of $\ddot{\xi}$ to obtain
\be
 \ddot{\xi}_{\text{full}} = \frac{2}{m^2} \Delta p_0^2 
 + C_{\text{rel}} + C_{\text{LGUP}} + C_{\text{QGUP}} + C_{\text{LGUP}}^{\text{rel}}, \label{masteq}
\ee
where
\ber
C_{\text{rel}} &=& -\frac{2}{m^4c^2} (\langle p_0^4\rangle - \langle p_0 \rangle \langle p_0^{3} \rangle)+ \frac{\Delta {p_0^{(3)}}^2}{2m^6c^4} \label{c1} \\
C_{\text{LGUP}} &=& -\frac{12\alpha}{m^2} (\langle p_0^3\rangle - \langle p_0 \rangle \langle p_0^{2} \rangle) + \frac{18\alpha^2}{m^2} {\Delta {p_0^{(2)}}^2} \label{c2} \\
C_{\text{QGUP}} &=& \frac{40\alpha^2}{m^2} (\langle p_0^4\rangle - \langle p_0 \rangle \langle p_0^{3} \rangle) +\frac{200\alpha^4}{m^2} {\Delta {p_0^{(3)}}^2} \label{c3} \\
C_{\text{LGUP}}^{\text{rel}} &=& \frac{2\alpha}{m^3c^2} (\langle p_0^5\rangle - \langle p_0 \rangle \langle p_0^{4} \rangle) +\frac{25\alpha^2}{2m^6c^4} {\Delta {p_0^{(4)}}^2}. \label{c4}
\eer

The master equation \eqref{masteq} has the following solution giving the rate of broadening of the free wavepacket under the combined influence of the relativistic and GUP corrections
\begin{widetext}
\be
\Delta x (t) = \sqrt{\xi_{\text{in}} + \dot{\xi}_{\text{in}} t + \frac{(\Delta p_0^2)_{\text{in}}}{m^2} t^2  + \frac{1}{2} \left( C_{\text{rel}} + C_{\text{LGUP}} + C_{\text{QGUP}} + C_{\text{LGUP}}^{\text{rel}}\right)t^2},
\label{freegup1}
\ee
\end{widetext}
where, the subscript ``$\text{in}$'' corresponds to the initial value of the various quantities, such as the initial width ($\sqrt{\xi_{\text{in}}}$), the initial rate of expansion  $\dot{\xi}_{\text{in}}$ and the initial variance of the canonical momentum $(\Delta p_0^2)_{\text{in}}$, and new corrections due to the relativistic and GUP effects appearing in \eqref{masteq}.

We now compute the expansion rates by considering a normalized Gaussian wave-packet of the form
\begin{equation*}
    \psi(x) = \frac{1}{(2\pi\xi)^{1/4}}\exp\left(\frac{i}{\hbar} \overline{p}_0 x - \frac{(x- \overline{x})^2}{4\xi}\right),
\end{equation*}
which represents a minimum wave-packet with $\langle x \rangle = \overline{x}$, $\langle p_0 \rangle = \overline{p}_0$, ${\Delta x}^2 = {\langle x^2 \rangle - \langle x \rangle^2} = \xi$, and $\Delta p_0 = \frac{\hbar}{2\sqrt{\xi}}$. Its Fourier transformation in momentum space is
\begin{eqnarray}
 \phi(p_0) &=& \frac{1}{\sqrt{2\pi\hbar}}\int_{-\infty}^{+\infty} \psi(x) e^{-ip_0x/\hbar} \nonumber \\
 &=& \left(\frac{2\xi}{\pi\hbar^2}\right)^{1/4}\exp\left({-\frac{ix_0}{\hbar}(p-\overline{p}_0) - \frac{(p-\overline{p}_0)^2 \xi}{\hbar^2}}\right)\nonumber
\end{eqnarray}

Since our results contain moments of $p$ upto the eighth order, and using the standard quantum mechanical definition $\langle p_0^n \rangle =\int_{-\infty}^{+\infty}\phi^*(p_0)p_0^n\phi(p_0)$ we calculated following coefficients for the gaussian wavepacket, 
\begin{eqnarray}
 C_{\text{rel}} &=& \frac{3 \hbar ^2}{128 c^4 m^6 \xi_{\text{in}}^3} \left(-16 c^2 m^2 \xi_{\text{in}} \left(4 \overline{p}_0^2 \xi_{\text{in}} + \hbar^2 \right) \right. \nonumber \\
 && \left. + 48 \overline{p}_0^2 \xi_{\text{in}} \hbar^2 + 48 \overline{p}_0^4 \xi_{\text{in}}^2 + 5 \hbar^4\right) \label{crgau}\\
 C_{\text{LGUP}} &=& \frac{3 \alpha  \hbar ^2 }{4 m^2 \xi_{\text{in}}^2} \left(3 \alpha  \hbar ^2+24 \alpha  \overline{p}_0^2 \xi_{\text{in}}-8 \overline{p}_0 \xi_{\text{in}}\right) \label{clgau}\\
 C_{\text{QGUP}} &=& \frac{15 \alpha ^2 \hbar ^2}{8 m^2 \xi_{\text{in}}^3} \left(25 \alpha^2 \hbar^4+16 \overline{p}_0^2 \left(15 \alpha ^2 \xi_{\text{in}} \hbar ^2+ \xi_{\text{in}}^2\right) \right. \nonumber \\
 && \left. +240 \alpha ^2 \overline{p}_0^4 \xi_{\text{in}}^2+4 \xi_{\text{in}} \hbar^2\right) \label{cqgau}\\
 C_{\text{LGUP}}^{\text{rel}} &=& \frac{\alpha  \hbar^2}{16 c^4 m^6 \xi_{\text{in}}^4} \left[8 c^2 m^3 \overline{p}_0 \xi_{\text{in}}^2 \left(4 \overline{p}_0^2 \xi_{\text{in}} +3 \hbar^2\right) \right. \nonumber\\
 && \left. +25 \alpha  \left(84 \overline{p}_0^4 \xi_{\text{in}}^2 \hbar^2+48 \overline{p}_0^2 \xi_{\text{in}} \hbar ^4 \right.\right. \nonumber\\
 && \left.\left. +32 \overline{p}_0^6 \xi_{\text{in}}^3+3 \hbar ^6\right)\right]. \label{clrgau}\\ \nonumber
\end{eqnarray}

As can be seen from (\ref{c1}-\ref{c4}), 
the above modifications contain moments up to the eighth order in momentum space.  
It is indeed a nontrivial result, as it
shows that not only the standard deviation, but also higher order moments such as the skewness ($3^{\text{rd}}$), kurtosis ($4^{\text{th}}$), hyperkurtosis ($5^{\text{th}}$), hypertailedness ($6^{\text{th}}$) etc. all dictate the broadening rate, albeit with decreasing importance. 
Equipped with the above, we ask our 
primary question of interest -
can the above GUP modifications be observed in an experiment, similar to earlier analyses 
where it was shown that 
a large parameter space of GUP can be probed by measuring the DTD for large molecular wavepackets such as $C_{60}$, $C_{176}$ \cite{carlos1, carlos2}? 
%
%
The corresponding results incorporating
relativistic effects are given by 
\eqref{crgau} - \eqref{clrgau}. It can be easily checked the results of \cite{carlos1, carlos2}
are recovered in the $c\rightarrow\infty$ limit.
%
%
As in the above references, we address
this question numerically.  

The DTD is defined as
\begin{equation}
  \Delta t_{\text{double}} =  t^{\text{GUP}}_{\text{double}} - t^{\text{HUP}}_{\text{double}},
  \label{dtd}
\end{equation}
where the first and second terms on the right hand side signify the 
times required for a free wavepacket to double its width following \eqref{freegup1}, and by 
the same equation in the $\alpha \rightarrow 0$
limit (i.e. no GUP). 
Note that even in the latter limit, 
the relativistic effect, in terms of 
$C_{\text{rel}}$ is always present in 
\eqref{freegup1}.


To calculate DTD using \eqref{freegup1}, we first notice that, since we are working with a Gaussian wavepacket, the term $\dot{\xi}_{\text{in}}=0$.
%
%
%
Next, one can replace initial value of
$\Delta p_0$ in terms of the initial position uncertainty $\Delta x$ 
in \eqref{freegup1}, 
using the following minimum uncertainty relation
(again since $[x,p_0]=i\hbar$)
\begin{eqnarray}
 (\D x)_{\text{in}} (\D p_0)_{\text{in}} = \frac{\hbar}{2} , 
\label{gunc}
\end{eqnarray}
and solve for 
the doubling time with GUP
in which the width becomes $2\Delta x_0$. To find the doubling time without GUP, we simply set the GUP parameter to
zero in the above result. 
This enables us to calculate the doubling time difference \eqref{dtd} which becomes a function of the initial width ($\Delta x_0$), mass ($m$), mean velocity ($v$) of the wavepacket, as well as, the Planck constant ($\hbar$), speed of light ($c$), and the value of the GUP parameter ($\alpha$).


We calculate the DTD numerically for three different molecular wavepackets; these are - (i) Buckyball $C_{60}$, (ii) Buckyball $C_{176}$ and (iii)  Tetraphenylporphyrin or TPPF152 molecule ($C_{168}H_{94}F_{152}O_{8}N_{4}S_{4}$). Relevant physical parameters for these molecular wavepackets are given in the accompanying table.
\begin{table}
\centering
  \begin{tabular}{ | c | c | c | c | } 
  \hline
  Molecules & Mass (kg) & Width (m) & Velocity (ms$^{-1}$)\\ 
  \hline
  $C_{60}$ & $1.1967 \times 10^{-24}$ & $7 \times 10^{-10}$ & $10^5$ \\ 
  \hline
  $C_{176}$ & $3.5070 \times 10^{-24}$ & $1.2 \times 10^{-9}$ & $10^5$\\ 
  \hline
  TPPF152 & $8.8174\times10^{-24}$ & $6\times 10^{-9}$ & $10^5$\\ 
  \hline
\end{tabular}
\caption{Physical parameters of the wavepackets under consideration. While mass and width of the wavepackets are known from experiments, mean velocity is assumed by us which provides measurable effects.}
\end{table}
It is important to note that the aforementioned systems behave quantum mechanically 
and are stable against decoherence, at least for their assumed widths, as shown for example by means of 
double-slit experiments \cite{dbs1}-\cite{dbs3}. 
Therefore the GUP 
applies to them and would affect the 
broadening rates of these wavepackets. 
\footnote{Although it has been claimed that 
the GUP needs to be applied cautiously for a composite system (such as one with many constituent atoms) \cite{ac}, we adopt the point of view that GUP would apply to the quantum system as a whole \cite{pik1,pik2,bdm}. In the end, it is for experiments to decide on its correctness.}

\begin{figure*}
    \centering
    \begin{subfigure}
        \centering
        \includegraphics[width=0.48\linewidth]{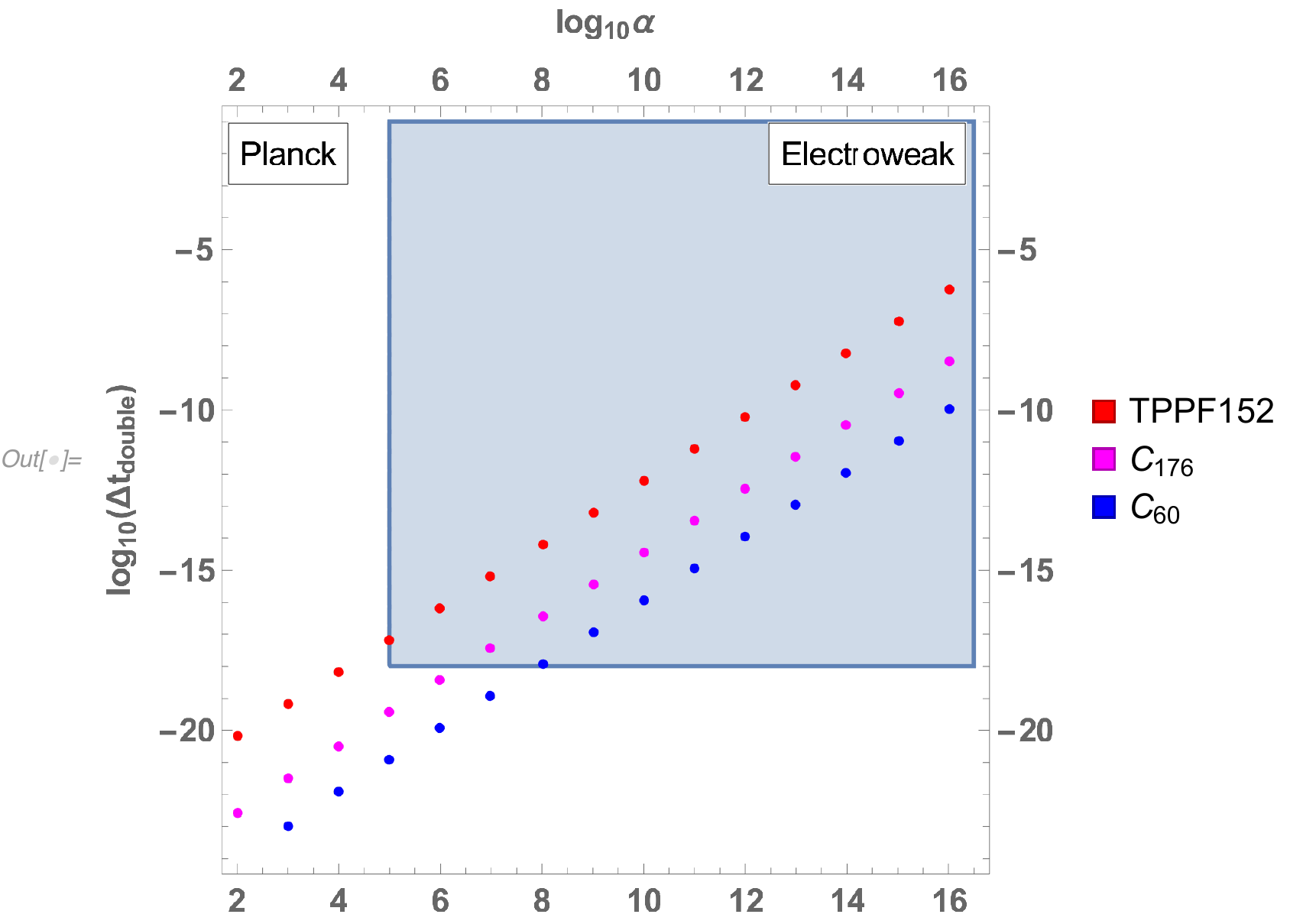}
    \end{subfigure}
    \begin{subfigure}
        \centering
        \includegraphics[width=0.48\linewidth]{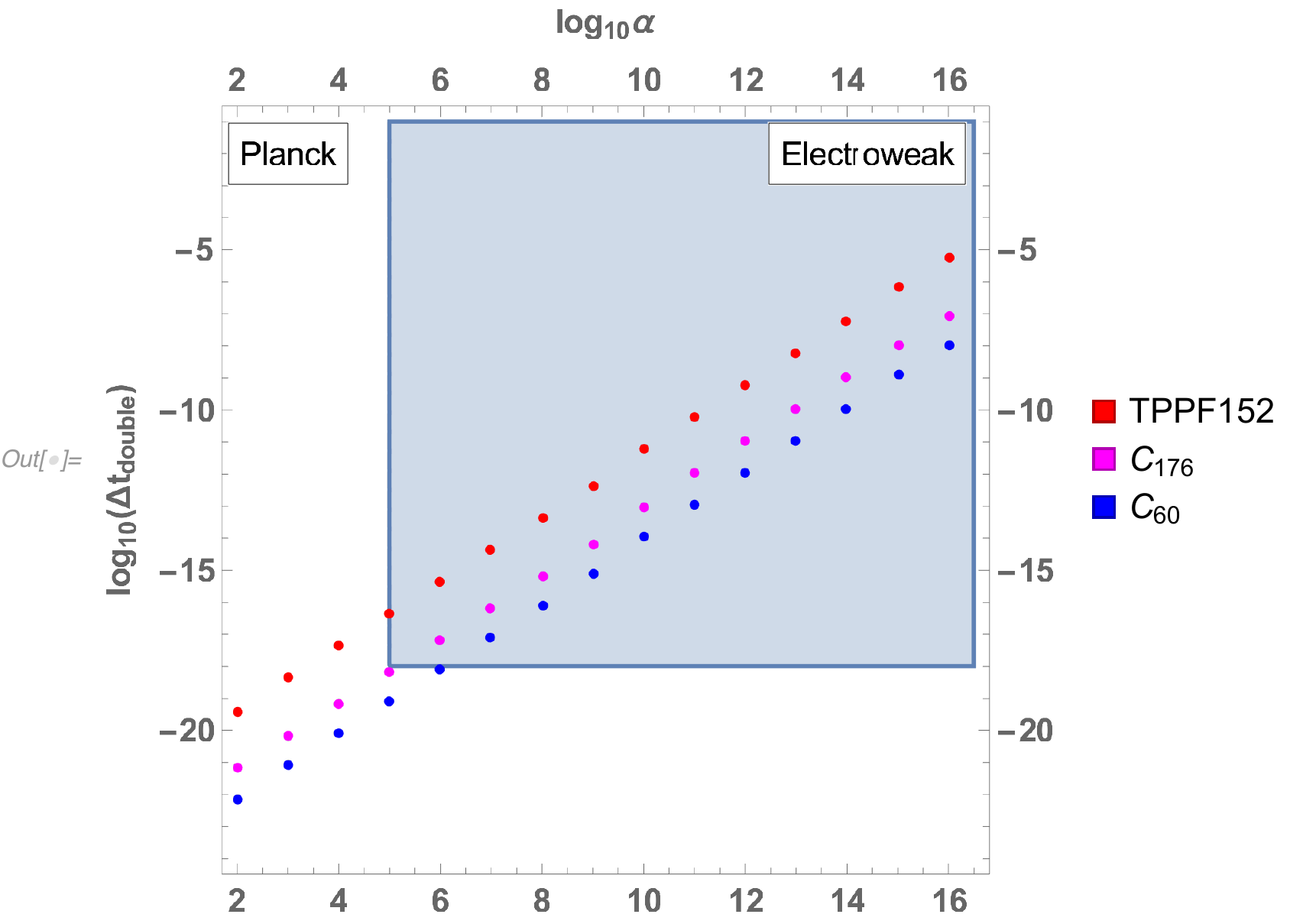}
    \end{subfigure}
    \caption{\label{fig} Doubling time difference versus GUP parameter plot in logarithmic scale. The left panel of the figure corresponds to the results with the relativistic correction, as carried out in the present paper, and the right panel is without considering the relativistic correction, as carried out in the earlier work \cite{carlos1}. The blue shaded regions correspond to the GUP parameter space that could be scanned by measuring DTD with attosecond ($10^{-18}$s) accuracy. The full GUP parameter space, up to the Planck scale could be scanned with zeptosecond accuracy ($10^{-21}$s). For details see text.}
\end{figure*}

Results of our numerical analysis are depicted in the left panel of Figure \ref{fig}, which is a log-log plot relating the GUP parameter $\alpha$ with the doubling time difference $\Delta t_{\text{double}}$. The plot covers the entire region between the electroweak scale and the Planck scale. The shaded blue region corresponds to the region of parameter space that can be probed by using these wavepackets and with an atomic clock with attosecond ($10^{-18}$ s) accuracy, which has already been achieved more than a decade ago (see, for instance \cite{atto}). 
With the $C_{60}$ molecule, we are expected to indirectly probe $8$ orders of magnitude up from the electroweak energy scale (and an equal order of magnitude down in the corresponding length scale).
With the $C_{176}$ we get an improvement of a further order of magnitude, while with TPPF152 one may be able to 
probe down to $11$ orders of magnitude away from electroweak scale. 
In other words, one should be able to 
probe the region of the parameter space $10^5 \le \alpha \le 10^{16}$ already with these wavepacket expansion experiments with an accuracy of attosecond timescale.
Further improvements are expected, 
with the advancement of atomic clock techniques and availability of larger wavepackets, with which one may 
be be able to probe reasonably close to the Planck scale itself. 
In fact, very recently the measurement of zeptosecond  time delay ($10^{-21}$s) was reported in \cite{zps}. 
As we can see from Fig. \ref{fig}, zeptosecond accuracy together with TPPF152 can probe all the way to the Planck scale, provided we can measure the DTD of this expanding wavepacket, moving with a velocity of $10^{5}$ m/s. This amount to scanning the full parameter space of (linear) GUP $1 \le \alpha \le 10^{16}$.  We know no other mechanism which can provide such an extraordinary and possibly complete scanning of the GUP/QG parameter space. 
Note that, because of the relationship between the GUP parameter $\alpha$ with the minimal length, via eq. (\ref{alpha}), this is equivalent of searching for the minimal length up to $10^5\, l_{\text{Planck}}$ with attosecond accuracy and up to the Planck length with zeptosecond accuracy.

Finally, we provide a comparison of the results with or without relativistic correction. To do this, we include a right panel in figure \ref{fig}, which is without the relativistic corrections, first carried out by one of us in an earlier work \cite{carlos1}. 
By comparison, we see that the relativistic corrections do change results for the doubling time difference for all of the three wavepackets and for the entire limit $1\le \alpha \le 10^{16}$. Understandably, the difference is more pronounced for larger values of $\alpha$. Also, as expected, the less massive wavepackets, such as $C_{60}$ and $C_{176}$ are strongly affected as compared with the heavy TPPF152 wavepacket. 


To conclude, the present study attempts to bridge the apparently formidable gap between QG theory and its potential verification by experiments. In particular, we have proposed to study wavepacket expansion experiments
with the hope of either seeing some of the predicted effects, or in their absence, imposing stringent bounds on QG parameters. In particular, we have considered the broadening of molecular wave packets for a set of well-studied large molecular systems moving at relatively high speeds, such that neither
relativistic nor QG effects in their evolution are insignificant. 
We computed the time taken for the corresponding wavepackets to double in size and the showed that QG/GUP effects entail a measurable difference in the doubling times, which may just be measurable with current precision of time-measurements, or those that are projected in the near future, as
clearly demonstrated in the accompanying figures!
Taking the required relativistic effects into account, we showed that the some of the earlier conclusions don't just remain, they in fact get further solidified. 
Again, the unprecedented accuracy of time measurements should aid in this measurement, which of course would get progressively even better in the future. Note that the detection of a $\alpha_0 \gg 1$ would signify a length scale intermediate between the electroweak and the Planck scale. Even if the 
predicted effects are not observed, that 
would provide the best constraints on the GUP parameter $\alpha$ to date. For example, with an attosecond accuracy, we would provide up to $5$ orders of magnitude tighter bound than previous best bound by measuring Lamb shift \cite{adv}. Furthermore, with the latest implementation of time measurement in the zeptosecond order, we would be able scan the whole GUP parameter space, and thus verifying or rulling out the linear GUP modification altogether!
We hope to continue our study of similar effects in other quantum systems that can be prepared in the laboratory and report elsewhere.


{\bf{Acknowledgement:}}
This work is supported by the Natural Sciences and Engineering Research Council of Canada. Research of SKM is supported by CONACyT research grant CB/2017-18/A1S-33440, Mexico.
%



\begin{thebibliography}{99}
%
\bibitem{carlos1} C. Villalpando and S.K. Modak, Class.Quant.Grav. 36 (2019) 21, 215016.
\bibitem{carlos2}
C.~Villalpando and S.~K.~Modak,
Phys. Rev. D \textbf{100}, no.2, 024054 (2019) [arXiv:1812.06112 [gr-qc]].


\bibitem{planck} M. Planck, Preuss. Akad. Wiss., S.479-480 (1899).

\bibitem{tomil} K.A. Tomilin, 
Proceedings Of The XXII Workshop On High Energy Physics And Field Theory. pp. 287–296, (1999).

\bibitem{hberg} W. Heisenberg, Ann. Phys. 5, 32 (1938).

\bibitem{yang} C. N. Yang, 
Phys. Rev. 72, 874 (1947).

\bibitem{deser} S. Deser, 
Reviews of Modern Physics, 29, 417-423 (1957).

\bibitem{maed1}C. A. Mead, 
Phys. Rev. 135, B849 (1964).

\bibitem{maed2}C.A. Mead, 
Phys. Rev. 143, 990 
(1966).

\bibitem{garay}
L. J. Garay,  
Int. J. Mod. Phys. {\bf A10}, 
10(02):145–165, 1995.

\bibitem{amo1} P. Kolorenc {\it et al.}, Phys. Rev. A 82, 013422 (2010).

\bibitem{amo2} F. Trinter {\it et al.}, Phys. Rev. Lett. 111, 093401 (2013).

\bibitem{gup1}
D. Amati, M. Ciafaloni, G. Veneziano,
Phys. Lett. {\bf B216} (1):41–47, 1989.

\bibitem{gup2}
D. J. Gross and P. F. Mende,
Nucl. Phys. {\bf B303} 407–454 (1988).

\bibitem{gup3}
C. Rovelli and L. Smolin,
Nucl. Phys {\bf B 442} 593–619 (1995).

\bibitem{kpp} K. Konishi, G. Paffuti, P. Provero, 
Phys. Lett. B 234, 276 (1990).

\bibitem{golam}G. M. Hossain, V. Hussain and S. S. Seahra, 
Class. Quant. Grav. 27, 165013 (2010).

\bibitem{gup4}
M. Maggiore,
Phys. Lett. {\bf B 304}, 65–69 (1993).

\bibitem{smolin} J. Magueijo, L. Smolin, 
Phys. Rev. Lett. {\bf 88}{}, 190403 (2002).

\bibitem{dv1}
S. Das, E. C. Vagenas,
Phys. Rev. Lett. {\bf 101}, 221301 (2008). 

\bibitem{adv}
A. Ali, S. Das, E. C. Vagenas, 
Phys. Rev. {\bf D84}, 044013 (2011). 

\bibitem{bdm}
P. Bosso, S. Das, R. B. Mann, 
Phys. Lett. {\bf B785}, 498-505 (2018).

\bibitem{doug}
M.~Bishop, E.~Aiken and D.~Singleton,
Phys. Rev. D \textbf{99}, no.2, 026012 (2019).

\bibitem{dbs1} Markus Arndt et. al., Nature, vol. 401, pp. 680 - 682 (1999).

\bibitem{dbs2} A. Goel, J. B. Howard and J. B. V. Sande, Carbon 42 1907-1915 (2004).

\bibitem{dbs3} S. Gerlich, S. Eibenberger, M. Tomandl, S. Nimmrichter, et. al., Nat. Commun. 2, (2011) 263.

\bibitem{ac}
G. Amelino-Camelia, 
Phys. Rev. Lett. {\bf 111}, 101301 (2013). 

\bibitem{pik1}
I. Pikovski, M. R. Vanner, M. Aspelmeyer, 
M. S. Kim, Časlav Brukner 
Nature Physics {\bf 8}, 393–397(2012). 


\bibitem{pik2}
P. Bosso, S. Das, I. Pikovski, M. R. Vanner,
Phys. Rev. {\bf A96}, 023849 (2017).


\bibitem{atto} S. Baker {\it et al.}, Science
Vol. 312, Issue 5772, pp. 424-427 (2006).

\bibitem{zps} S. Grundmann {\it et al.}, Science Vol. 370, Issue 6514, pp. 339-341 (2020).

\end{thebibliography}
\end{document}